\documentclass[aip,reprint,twocolumn,amsmath,amssymb]{revtex4-2}
\usepackage[T1]{fontenc}
\usepackage[utf8]{inputenc}
\usepackage{graphicx}
\usepackage{color}
\usepackage{babel}

\begin{document}
\title{Comment on ``Impact of particle number and cell-size in fully implicit
charge- and energy-conserving particle-in-cell schemes'' by N. Savard
et al., Phys. Plasmas 32, 073903 (2025)}
\author{L. Chacón}
\email{chacon@lanl.gov}

\author{G. Chen}
\affiliation{Los Alamos National Laboratory, Los Alamos, NM 87544}

\author{L. Ricketson}
\affiliation{Lawrence Livermore National Laboratory, Livermore, CA 94550}

\maketitle

\section{Introduction}
We take issue with the conclusions in the recent publication by Savard
et al.\citep{savard2025impact} In the reference, the authors implement
a fully nonlinear charge- and energy-conserving implicit particle-in-cell
method (ECC-IPIC\citep{chacon2013charge,chacon2016curvilinear}),
and use it to study the impact of particle number in the quality of
the ECC-IPIC solutions for several problems, including an ion acoustic
shock-wave (IASW\citep{chacon2013charge}) problem
and several sheath problems in bounded plasmas. 

From the study, the authors concluded that ``to reproduce highly resolved
convergent solutions, a higher amount of particles per cell need to
be used in the implicit scheme for both periodic and bounded simulations
when the cell size exceeds the Debye length'' (quote from the abstract
in Ref.\ \onlinecite{savard2025impact}). We demonstrate in the sequel that,
according to our analysis for the IASW test, this conclusion does not survive independent
scrutiny. We have identified several diagnostics procedural issues
that are at the root of their conclusion, which when fixed dramatically
change the outcome of the study.

It is not our goal to study all the different variants of the ECC-IPIC
method that the authors considered in their study, nor to look at all
the different numerical examples. It is also not our goal to demonstrate speedups of ECC-IPIC vs.\ explicit PIC, since these are documented elsewhere,\cite{chacon2013charge} and Savard et al.\ also report them (a speedup of $5\times$ of adaptive-mesh ECC-IPIC vs. explicit PIC is reported in Table I\citep{savard2025impact}). Since the authors have chosen
to document a negative result (that ECC-IPIC algorithms are less accurate
than their explicit counterparts when underresolving on particles), which stems from methodological issues in their analysis, 
it should be sufficient to focus on one representative example using one ECC-IPIC
variant to show that the stated conclusions do not hold due to key shortcomings of their analysis.

\begin{figure}
\begin{centering}
\includegraphics[viewport=0bp 0bp 540bp 432bp,clip,width=1.0\columnwidth]{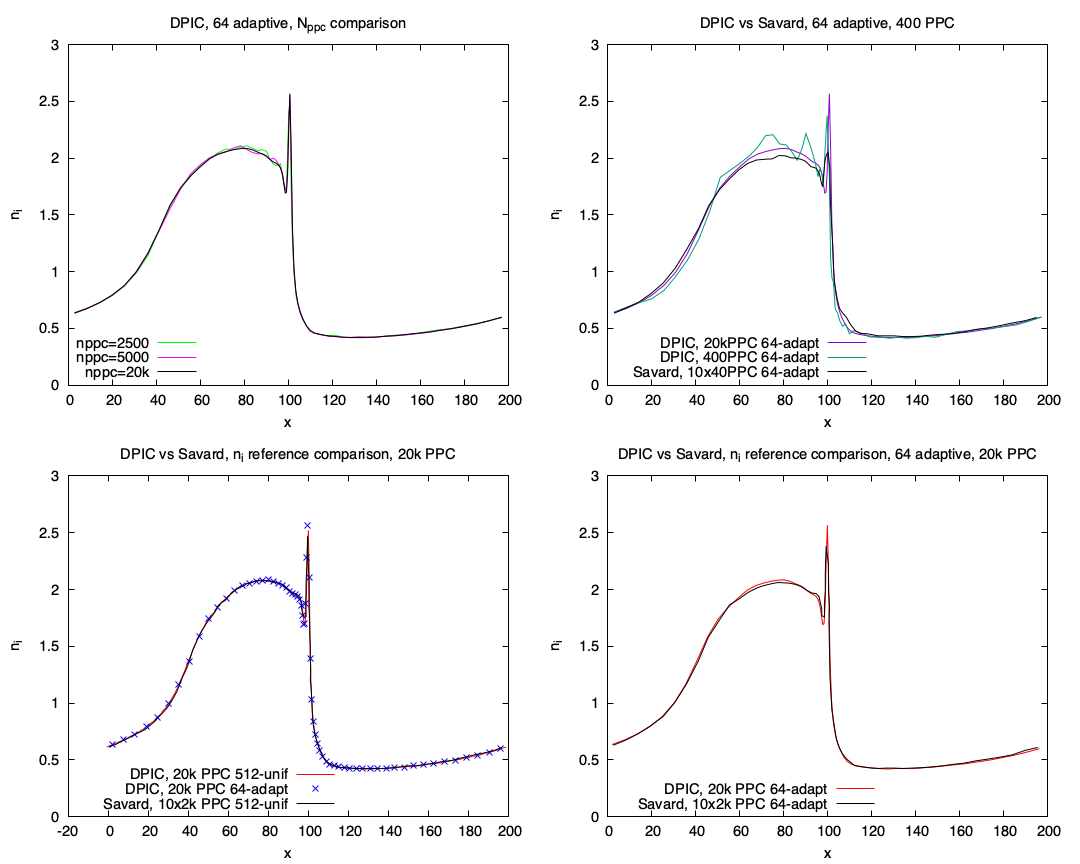}
\par\end{centering}
\caption{Comparison of well-resolved IASW final solutions on both uniform
and adaptive meshes from our ECC-IPIC implementation (DPIC) as well
as Savard's (from Ref.~\onlinecite{savard2025impact}).}\label{fig:well_resolved_comp}

\end{figure}
With this in mind, we choose the IASW problem to make our case, as it is a benchmark
we know very well.\citep{chacon2013charge} The IASW is a traveling shock-wave problem simulated in the shock moving frame with periodic boundary conditions. It is a closed system, with linear momentum and total energy strictly conserved. In Ref. \onlinecite{savard2025impact},
the authors adequately document the formulation and implementation
of their ECC-IPIC algorithm. As Savard, we evolve the electrostatic potential by taking the divergence of Ampere's
equation (Eq.\ 1 of Ref. \onlinecite{savard2025impact}).
We employ the same mesh setup [with the same variable logical-to-physical-coordinate
map $x(\xi)$], same timestep ($dt=2\omega_{pe}^{-1}$) and initial conditions (except for particle initialization, which Ref.\ \onlinecite{savard2025impact} did not provide),
and the so-called (by Savard et al.) ``implicit cloud-in-cell'' (ICIC) gather-scatter strategy
as described in the reference. So, to our knowledge, our discrete
formulation is equivalent to theirs, and in fact our well-resolved
solutions in both adaptive and uniform meshes look equivalent (see
Fig.~\ref{fig:well_resolved_comp}). 

We focus our attention instead on the particle initialization (pre-simulation) and the simulation diagnostics (post-simulation). 
While the
ability to use mapped meshes in PIC allows distributing mesh degrees of freedom efficiently, that capability requires utmost care in the implementation of both steps. 
As we argue below, lack of attention to these is at the root of the observations by Savard et al.

\section{Particle initialization}

Savard et al. do not really specify how they initialize particles
for their study, other than to say that the number of particles per
cell (PPC or $N_{ppc}$) ``is an average quantity over all cells''
(page 073903-6, bottom of second column in Ref. \onlinecite{savard2025impact}).
They do not indicate how the particle number is chosen spatially, or
how the initial Maxwellian distributions are sampled in velocity space.
Therefore, we are unable to reproduce their simulations exactly.
We surmise that they likely used a noisy Box-Muller random sampling algorithm\cite{box1958note} for
the initial Maxwellians, but their discussion does not make clear
how the particles are distributed in space. 

Since particle noise in the density correlates with the number of
particles per cell,\citep{tranquilli2022deterministic} and this corresponds to uniform noise in the logical space density $Jn$, a natural
choice to keep noise uniformly distributed is to distribute particles
uniformly in logical space (i.e., with all logical cells having the
same number of particles, rather than ``on average''). However,
with that choice, in adaptive-mesh PIC simulations one must vary the
particle weight on a per-cell basis to match the initial density profile.
This variable weight initialization can be performed approximately
by setting the particle weight at cell $i$ to be $w_{p,i}\propto(Jn)_{i}/N_{ppc}$
(with $J_{i}=\Delta x_{i}/\Delta\xi$ the Jacobian factor of the curvilinear
map at cell $i$). This will result in a noisy initialization of the
density profile. However, one can do better and match the density
profile exactly by solving a mass matrix problem.\citep{chen2021unsupervised}
We consider both approaches here. 

The mass-matrix approach postulates a particle weight per cell $W_{i}$,
such that any particle belonging to cell $i$ inherits that weight,
i.e., 
\begin{equation}
w_{p}=\sum_{j}W_{j}S_{0}(x_{j}-x_{p}),\label{eq:npg_w_p}
\end{equation}
with $S_{0}$ the zeroth-order (nearest-grid-point or top-hat) B-spline interpolation
kernel. From the density gather expression for ICIC, introducing Eq.\
\ref{eq:npg_w_p} we find:
\begin{align}
(Jn)_{i}=\frac{1}{\Delta \xi} \sum_{p}w_{p}S_{2}(x_{i}-x_{p}) &= \\
\sum_j\underbrace{\frac{1}{\Delta \xi}\sum_{p}S_{2}(x_{i}-x_{p})S_{0}(x_{j}-x_{p})}_{M_{ij}}W_{j} &= \sum_j M_{ij}W_{j},
\end{align}
from which $\{W_{j}\}$ can be found by inversion of the matrix $\{M_{ij}\}$,
and the particle weights $w_{p}$ follow from Eq. \ref{eq:npg_w_p}.
The cost of forming this matrix is linear with the number of particles, and we have found it in practice to be very well conditioned and easy to invert iteratively
to machine precision. This procedure ensures that our initial condition
matches the analytically specified one at cell centers to machine precision.

Regarding particle sampling, we consider various sampling approaches in both the configuration and the velocity space to understand their impact. In particular, we consider both random (RNG) and quiet start (QS). For QS, we use the Hammersley quasi-random low-discrepancy
series,\citep{hammersley1960monte} which features better convergence properties than standard Monte
Carlo.\cite{caflisch1995modified}

\section{Diagnostics}

Regardless of the initialization details, we have found that the most important factor
 determining the poor accuracy outcomes reported for implicit PIC in Ref.\ \onlinecite{savard2025impact}
is the diagnostic procedure. In the reference, the authors
perform ten runs and ensemble-average the results. The
reason for this choice is not stated explicitly, although it seems
 intended to reduce statistical noise. 

Our results support the hypothesis that this ensemble-averaging procedure
is likely at the root of the solution degradation observed by Savard
et al. To demonstrate this, instead of ensemble-averaging over 10
runs, we perform one simulation with $10\times$ the number of particles
per cell specified by Savard et al. For a given mesh, the computational
cost of both approaches is approximately the same, so the only remaining
discriminating factor to select either approach is accuracy. Because
 ECC-IPIC is nonlinear, the external ensemble
averaging operator does not commute with the ECC-IPIC solution propagator,
and the two approaches are therefore not equivalent. As we will show next, running
with $10\times$ more particles leads to much more acceptable accuracy
outcomes for ECC-IPIC simulations for similar computational cost than
reported by Savard et al. We will also identify the key reason behind
the degradation of the ensemble-averaged diagnostic.

We first note that the error metric proposed by Savard et al. (Table
I of Ref. \onlinecite{savard2025impact}),
\begin{equation}
\sigma_{ion,RMS}=\sqrt{\frac{1}{N_{x}}\sum_{i}\left(\frac{n_{ion}-n_{ref}}{n_{ref}}\right)_{i}^{2}},\label{eq:savard_rms}
\end{equation}
is problematic in several ways. Here, $i$ represents cell centers
and $n$ ion densities, and $n_{ref}$ is the reference solution.
In Savard's study, the density reference is taken as the uniform 512-mesh
ECC-IPIC solution. Using the uniform-mesh solution as a reference
conflates two sources of error; namely, the spatial discretization and the particle
noise (which, as we will see, become comparable for sufficiently large
PPC for the IASW), and may obscure the true PPC convergence properties
of ECC-IPIC. Additionally, the definition in Eq.\ \ref{eq:savard_rms}
may amplify the true error because it does not account for the fact that
the ECC-IPIC shock solution may undergo a slight phase shift due to imperfect momentum
conservation (unlike explicit PIC, which is strictly momentum conserving,
and does not have this issue). Given the sharpness of the IASW shock
front at the final time, even small shifts can lead to large
pointwise discrepancies, obscuring the real convergence properties of the scheme.\footnote{In fact, numerical error of shocked solutions scales as $\mathcal{O}(\Delta x^{1/p})$, where $\Delta x$ is the grid spacing, and $p$ defines the error norm used, e.g., $p=1,2,\infty$ for absolute-value, Euclidean, and maximum norm, respectively. Note that this is in stark contrast to smooth functions, for which $\mathcal{O}(\Delta x)$ shifts give rise to $\mathcal{O}(\Delta x)$ errors \textit{regardless} of the norm being used.} 

To isolate particle error effects, we use an adaptive-mesh simulation
with 20000 PPC as the reference, in place of Savard's $10\times2000$
PPC uniform-mesh reference. To separate errors due to possible solution shifts, we minimize
the RMS error over a range of phase-shifts $h$:
\begin{equation}
\small
\sigma_{ion,RMS}^{opt}=\min_{h\in(-h_{0},h_{0})}\sqrt{\frac{1}{N_{x}}\sum_{i}\left(\frac{n_{ion}(\xi_{i}+h)-n_{ref}(\xi_{i})}{n_{ref}(\xi_{i})}\right)^{2}},\label{eq:opt_sigma_rms}
\end{equation}
with $h_{0}=L_{x}/100$, and we also track the phase-shift that attains the minimum, $h_{opt}$, as a function of
PPC. The density profile $n_{ion}$ in Eq. \ref{eq:opt_sigma_rms} is evaluated at
$(\xi_{i}+h)$ using linear interpolation.

\begin{figure}
\begin{centering}
\includegraphics[width=1.0\columnwidth]{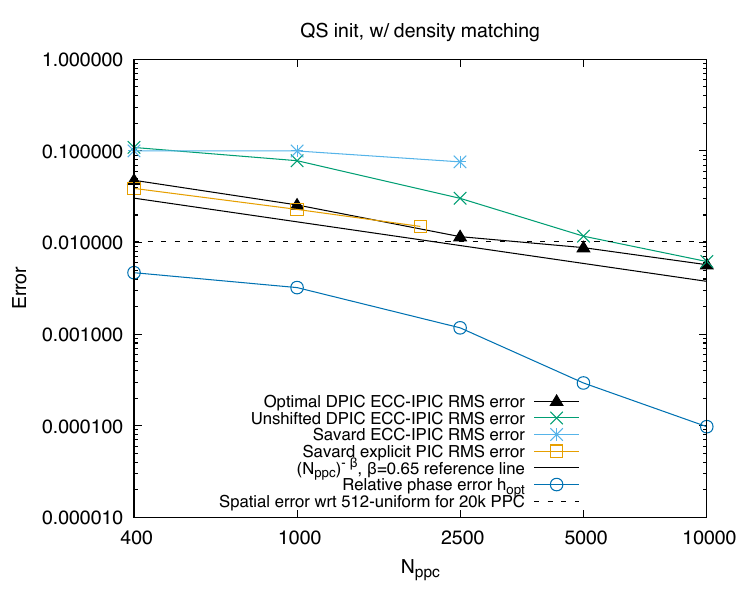}
\includegraphics[width=1.0\columnwidth]{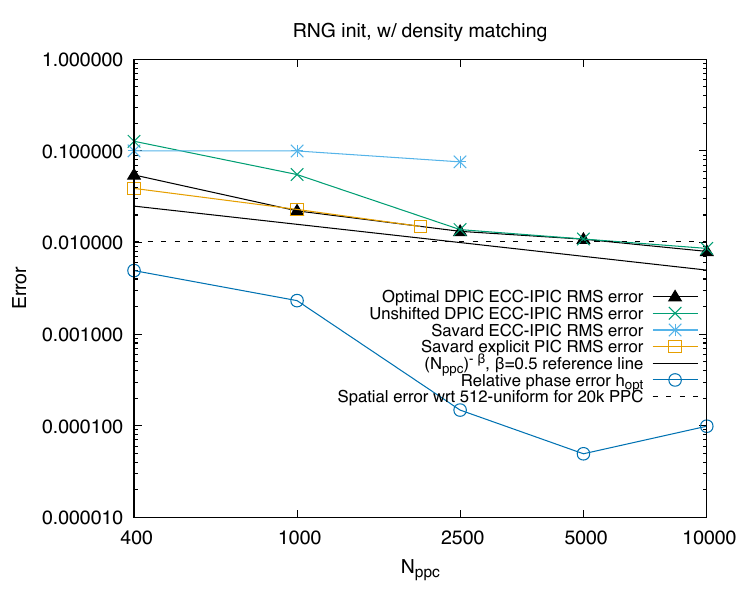}
\includegraphics[width=1.0\columnwidth]{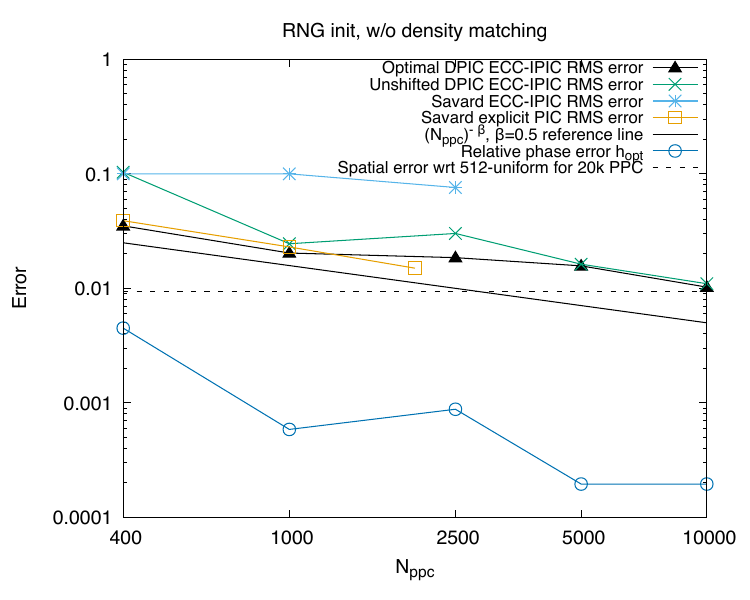}
\par\end{centering}
\caption{Error analysis of the ECC-IPIC algorithm with respect to the number
of particles, $N_{ppc}$. Top: quiet-start (QS) initialization with density matching. Middle: random initialization (RNG) with density matching. Bottom: RNG initialization without density matching. See text for description of various curves. Except for the spatial error level line, the reference solution is taken as the adaptive-mesh
simulation with 20k PPC.}\label{fig:error_analysis}
\end{figure}

\section{Results}
Our main results are depicted in Fig.\ \ref{fig:error_analysis}, where we consider several initialization strategies: QS with density matching, RNG with density matching, and RNG without density matching. In the figures, we plot (i) the RMS error $\sigma_{ion,RMS}$ from Eq. \ref{eq:opt_sigma_rms}, (ii) 
the RMS error using Savard's original formula in Eq. \ref{eq:savard_rms}
for the \emph{same} data, (iii) the explicit and implicit PIC RMS errors documented by
Savard\citep{savard2025impact} in Table I, and (iv) the phase-shift
error $h_{opt}$. We also include a reference scaling for particle
error with $N_{ppc}$, and the spatial error floor in our adaptive-mesh ECC-IPIC simulation, assessed by comparing
the 20k-PPC adaptive-mesh result with the uniform-mesh counterpart. The spatial
error is computed by linearly interpolating the 20k-PPC 512-cell uniform-mesh ECC-IPIC
solution and minimizing the RMS vs. the adaptive-mesh solution
with respect to a phase shift, as in Eq. \ref{eq:opt_sigma_rms}.
Several observations follow from Fig. \ref{fig:error_analysis}:
\begin{enumerate}
\item The most impactful initialization feature on the quality of the convergence data appears to be exact density matching, which results in smaller RMS errors and cleaner  PPC scalings. Nevertheless, all initializations considered provide qualitatively similar outcomes, inconsistent with Savard's conclusions.
\item $\sigma_{ion,RMS}^{opt}$ is commensurate with the explicit RMS error reported by Savard
et al., calling into question their claims that ECC-IPIC does not achieve  accuracy comparable to explicit PIC under the conditions considered.
\item The phase-shift error $h_{opt}$ is small but not negligible for the
coarser particle resolutions, and converges with $N_{ppc}$ similarly to the RMS error. Convergence of $h_{opt}$
accelerates significantly for larger PPC with density matching. Note that the presence of a phase-shift
error is particularly deleterious when ensemble-averaging
ECC-IPIC runs, as the phase shift will be different per run, and
the averaging tends to smooth the shock profile and introduce
a systematic bias. Evidence of such ensemble-averaging bias
is shown in Fig.\ \ref{fig:dpic_vs_savard}, where we compare the
shock density profiles using the adaptive mesh for the 20k PPC reference, the 400PPC run, and the ensemble-averaged $10\times40$ PPC result reported by Savard et al. Savard's ensemble-averaged profile exhibits a visibly smoothed shock front and a downstream density (in $x\in(60,100)$) that is systematically below the reference solution, whereas DPIC's shock front remains sharp and the downstream solution oscillates around the reference one.
\begin{figure}
\begin{centering}
\includegraphics[viewport=540bp 432bp 1080bp 864bp,clip,width=1.0\columnwidth]{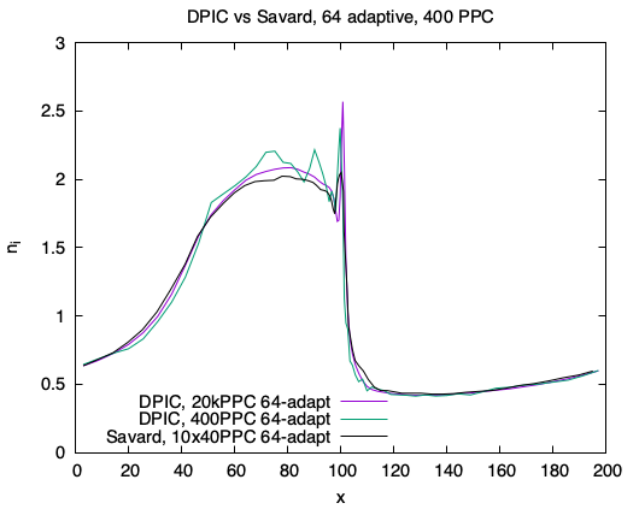}
\par\end{centering}
\caption{Comparison of particle-starved IASW solutions on adaptive meshes from
DPIC (400 PPC) and Savard's study ($10\times40$ PPC) vs the 20k PPC reference. Savard's ensemble-averaged solution
(black) shows systematic smoothing bias, whereas
DPIC's (green) does not.}\label{fig:dpic_vs_savard}
\end{figure}
\item Separating the shift error in Eq. \ref{eq:opt_sigma_rms} has a significant impact on the RMS accuracy. For QS, $\sigma_{ion,RMS}^{opt}$ scales as $N^{-0.65}$, faster than the
usual Monte Carlo (MC) scaling of $N^{-0.5}$ as a consequence of using low-discrepancy quasi-random sampling in our particle initialization. RNG recovers the standard MC scaling. The unshifted RMS error is significantly larger in magnitude than the optimized metric, and its decay with
$N_{ppc}$ closely follows the trend of phase shift error for all initializations, obscuring the true convergence behavior. 
\item Savard's
RMS error metric for ECC-IPIC does not exhibit the expected $N_{ppc}^{-0.5}$ Monte Carlo scaling, suggesting the presence of systematic errors that we attribute in part to ensemble-average-induced smoothing. Moreover, because ECC-IPIC advances the solution through a nonlinear implicit solve with finite tolerances, run-to-run variability in nonlinear convergence of multiple under-resolved implicit solves can introduce systematic (non-Monte-Carlo) error that ensemble averaging does not remove.
\item For density-matched setups, the spatial error introduced by the adaptive mesh exceeds the particle error
for large $N_{ppc}$, highlighting the need to use a reference solution on
the same spatial mesh to isolate particle errors from spatial discretization ones. 
\end{enumerate}

\section{Conclusions}

We have investigated the origin of the poor particle-number convergence
results of charge- and energy-conserving implicit PIC (ECC-IPIC) methods
on adaptive meshes reported by Savard and coauthors.\citep{savard2025impact}
Our analysis disagrees with the conclusion in the reference that ECC-IPIC
methods are less accurate than explicit PIC ones, and in fact demonstrates
that, when initialized and diagnosed carefully, ECC-IPIC schemes on
adaptive meshes produce accuracy levels comparable to their explicit
counterparts for equivalent number of particles per cell. Key improvements
needed to clean PPC convergence rates in ECC-IPIC include: 1) avoiding ensemble-averaging solutions
and favoring a single run with as many cumulative particles, and 2)
optimization of the RMS error by accounting for small phase shifts due to lack of strict momentum conservation. The observed lack of convergence in Ref.\ \onlinecite{savard2025impact} originates not from the ECC-IPIC algorithm itself, but from their diagnostic procedure, and is therefore largely preventable.
While our analysis focuses on only one of the numerical examples considered
by Savard et al. (IASW) and on one of the ECC-IPIC implementations (ICIC), it is sufficient to cast doubt on the generality of their conclusions.

\begin{acknowledgments}
This work was performed under the auspices of the U.S. Department of Energy (DOE) by  LANL and LLNL under contracts 89233218CNA000001 and DE-AC52-06NA25396, respectively, and supported by the DOE Office of Applied Scientific Computing Research.
\end{acknowledgments}

\bibliography{refs}

\end{document}